\documentclass[aps,prb,twocolumn,groupedaddress]{revtex4}

\usepackage[dvips]{graphicx}
\usepackage{dcolumn}
\usepackage{bm}
\usepackage{amssymb}
\usepackage{epstopdf}

\begin{document}


\title{
Incommensurate spin correlations induced by magnetic Fe ions substituted into overdoped Bi$_{1.75}$Pb$_{0.35}$Sr$_{1.90}$CuO$_{6+z}$


}


\author{H.~Hiraka$^{1}$, 
Y.~Hayashi$^{1}$,
S.~Wakimoto$^{2}$,
M.~Takeda$^{2}$,
K.~Kakurai$^{2}$,
T.~Adachi$^{3}$,
Y.~Koike$^{3}$,
I.~Yamada$^{4}$,
M.~Miyazaki$^{5}$,
M.~Hiraishi$^{5}$,
S.~Takeshita$^{6}$,
A.~Kohda$^{5,6}$,
R.~Kadono$^{5,6}$,
J.~M.~Tranquada$^{7}$,
K.~Yamada$^{8}$
}
\affiliation{$^{1}$Institute for Materials Research, Tohoku University, Sendai~980-8577, Japan}
\affiliation{$^{2}$Quantum Beam Science Directorate, Japan Atomic Energy Agency, Tokai, Ibaraki 319-1195, Japan}
\affiliation{$^{3}$Department of Applied Physics, Tohoku University, Sendai~980-8578, Japan}
\affiliation{$^{4}$Graduate School of Science and Engineering, Ehime University, Matsuyama~790-8577, Japan}
\affiliation{$^{5}$Department of Materials Structure Science, The Graduate University for Advanced Studies, Tsukuba, Ibaraki 305-0801, Japan}
\affiliation{$^{6}$Institute for Materials Structure Science, High Energy Accelerator Research Organization, Tsukuba, Ibaraki  3005-0801, Japan}
\affiliation{$^{7}$Condensed Matter Physics \& Materials Science Department, Brookhaven National Laboratory, Upton, New York 11973-5000, USA}
\affiliation{$^{8}$WPI Research Center, Advanced Institute for Materials Research, Tohoku University, Sendai~980-8577, Japan}


\date{\today}

\begin{abstract}

Spin correlations in the overdoped region of Bi$_{1.75}$Pb$_{0.35}$Sr$_{1.90}$CuO$_{6+z}$ have been explored with Fe-doped single crystals characterized by neutron scattering, muon-spin-rotation ($\mu$SR) spectroscopy, and magnetic susceptibility measurements.
Static incommensurate spin correlations induced by the Fe spins are revealed by elastic neutron scattering.
The resultant incommensurability $\delta$ is unexpectedly large ($\sim$0.2~r.l.u.), as compared with $\delta\sim 1/8$ in overdoped superconductor La$_{2-x}$Sr$_x$CuO$_4$.
Intriguingly, the large $\delta$ in this overdoped region is close to the hole concentration $p$.  This result is reminiscent of the $\delta\approx p$ trend observed in underdoped La$_{2-x}$Sr$_x$CuO$_4$; however, it is inconsistent with the saturation of $\delta$ in the latter compound in the overdoped regime.
While our findings in Fe-doped Bi$_{1.75}$Pb$_{0.35}$Sr$_{1.90}$CuO$_{6+z}$ support the commonality of incommensurate spin correlations in high-$T_{\rm c}$ cuprate superconductors, they also suggest that the magnetic response might be dominated by a distinct mechanism in the overdoped region.

\end{abstract}


\maketitle

\section{Introduction}


Neutron scattering studies have provided valuable characterizations of the momentum and energy dependences of spin correlations in high-$T_{\rm c}$ cuprate superconductors.
Incommensurate spin correlations (ISCs), which had been discovered in the early stage of high-$T_{\rm c}$ research,~\cite{yoshizawa88,birgeneau89} are one of the salient features of hole-carrier doped cuprates. 
The energy evolution of the ISCs in La$_{2-x}$Sr$_x$CuO$_4$ (La214),~\cite{tranquada04,vignolle07} YBa$_2$Cu$_3$O$_{6+y}$ (Y123),~\cite{stock05,hinkov07} and Bi$_2$Sr$_2$CaCu$_2$O$_{8+\delta}$ (Bi2212),~\cite{fauque07,guanguong09} shows a commonality of magnetic excitations in the form of an hourglass-like dispersion.
In the underdoped region, the incommensurability $\delta$ of static, as well as low-energy, ISCs is close to the hole concentration $p$ in La214~\cite{yamada98,matsuda00,fujita02} and Y123.~\cite{dai01}
As spin stripes with $\delta\simeq1/8$ are observed in Nd co-substituted La214 with $x=0.12$~\cite{tranquada95}, the linear relationship $\delta = p$ strongly suggests the presence of stripe correlations in the CuO$_2$ planes.
Another linearity between the onset $T_{\rm c}$ and $\delta$, $T_{\rm c}^{\rm on} = \delta$,~\cite{yamada98} as well as the development of the spin-gap below $T_{\rm c}$~\cite{yamada95,lake99} evinces an important impact of stripe correlations on the superconductivity.

In contrast, in the overdoped region, observation of ISCs becomes rather difficult because of peak broadening and doping-induced suppression of magnetic intensity, both of which probably result from a change in the degree of itinerancy of the electrons. 
Only slowly fluctuating ISCs are observable and $\delta$ remains nearly constant at around 1/8 in overdoped La214.~\cite{yamada98,lee03,wakimoto04}
However, because of the difficulty involved in synthesizing large single crystals, systematic neutron-scattering data on ISCs in the overdoped region have been restricted to the La214 system.


Recently, we succeeded in growing sizable single crystals of Bi$_{1.75}$Pb$_{0.35}$Sr$_{1.90}$CuO$_{6+z}$ [(Bi,Pb)2201].
With these crystals, the overdoped region is easily accessible over a wide range of $p$.
Here we report a search for spin correlations in overdoped (Bi,Pb)2201, where no well-defined spin correlations have been observed to date.~\cite{russo07}
In addition to the pristine sample, we also study samples with varying amounts of Fe substituted for Cu, as a strong effect of these magnetic dopants on spin correlations is expected, based on recent results in the La214 system.~\cite{fujita09,fujita09b} 
We find that quasistatic ISCs are induced by the large localized spins of the Fe dopants.
This observation for Bi-based cuprates supports an intrinsic nature of dynamic ISCs in high-$T_{\rm c}$ cuprates.
The obtained $\delta$ of $\sim0.2$ r.l.u., which is the largest value reported in any high-$T_{\rm c}$ cuprate, falls on the line $\delta \approx p$, in contrast to overdoped La214.
Differences in the nature of the spin correlations compared to the underdoped regime seem likely.

The rest of the paper is organized as follows.  In the next section, we describe the crystal growth and characterization by magnetic susceptibility, zero-field $\mu$SR, and elastic neutron scattering; the experimental results are also presented.  In Sec.~III, the implications for the nature of the magnetic correlations in overdoped (Bi,Pb)2201 are discussed.  The paper is summarized in Sec.~IV.

\section{Experiments and Results}

\subsection{Crystal growth and characterization}

Single crystals of Bi$_{1.75}$Pb$_{0.35}$Sr$_{1.90}$Cu$_{1-y}$Fe$_{y}$O$_{6+z}$ 
($y=0, 0.03, 0.06, 0.09$, and 0.13) 
were grown in air by the traveling-solvent floating-zone technique.
The amounts of metal elements determined by ICP spectrometry were in agreement with the nominal values to at least 95\%\ accuracy, except for $y=0.13(2)$ where the mol ratio between (Bi,Pb,Sr) and (Cu,Fe) changed from $4:1$ to $3.8:1.2$.
The crystal structure analyzed by X-ray-powder diffraction at room temperature was consistent with the \textit{Pnan} orthorhombic symmetry~\cite{torardi91} for all samples, and no significant peak broadening was observed even for $y=0.13$.
The $c$ lattice parameter decreased rapidly upon Fe doping, and the rate of decrease follows that observed for the crystalline Bi$_{1.8}$Pb$_{0.2}$Sr$_{2}$Cu$_{1-y}$Fe$_{y}$O$_{6+z}$,~\cite{xu00} thus evincing the Fe-atom doping from a structural viewpoint.
We also confirmed with neutron diffraction that incommensurate structural modulations along the $b$ axis characteristic to Bi-based cuprates are absent in our Pb$_{0.35}$-substitution samples, in accordance with previous STM studies.~\cite{nishizaki07}

$T_{\rm c}^{\rm on}$ was determined from magnetic shielding measurements.
The pristine samples show superconductivity below $T_{\rm c}^{\rm on} = 6$~K and $23$~K for as-grown and Ar-annealed ($600~^{\circ}$C $\times$ 5 days) samples, respectively.
For all the Fe-doped samples, on the other hand, no diamagnetism was detected down to 2~K.

\subsection{Magnetic susceptibility}

\begin{figure}[t] 
\centering
   \includegraphics[width=7.7cm]{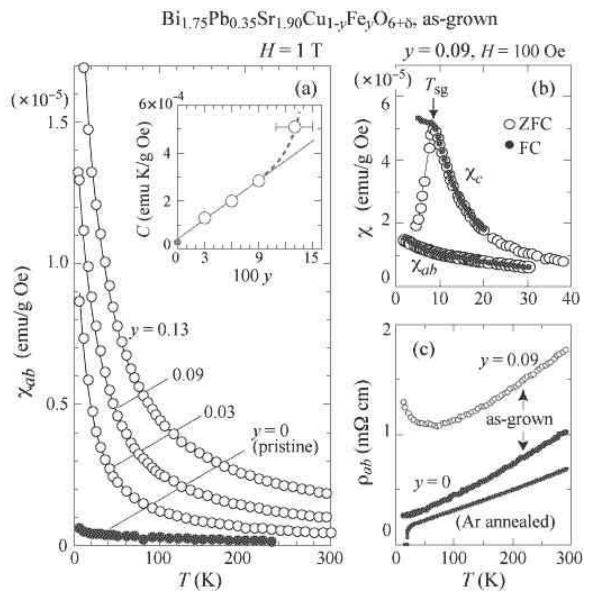}
   \caption{
   (a) $\chi_{ab}$ over a wide range of temperature for $H=1$~T. The inset shows $C$ as a function of the Fe content.
 (b) $\chi_{ab}$ and $\chi_{c}$ at low temperatures for $H=100$~Oe and $y=0.09$.
 (c) $\rho_{ab}$ for $y=0.09$ (as-grown) and $y=0$ (as-grown and Ar-annealed). }
   \label{imp-dep}
\end{figure}

Figure~\ref{imp-dep}(a) shows $\chi_{ab}(T)$ of the series of Fe-doped samples, in which a magnetic field of  $H=1$~T was applied in a direction parallel to the CuO$_2$ planes.
The Curie-Weiss law for high-temperature susceptibility was used to calculate the Curie constant~$C$ and the Weiss temperature ($|\Theta|\sim10$~K).
As can be seen in the inset of Fig.~\ref{imp-dep}(a), $C$ linearly increases up to $y=0.09$.
The effective number of Bohr magnetons~$p_{\rm eff}$ was found to be 4.4 using the equation $C=(N\mu_{\rm B}^2/3k_{\rm B})p_{\rm eff}^2$, by assuming that only the Fe spins contribute to $C$.
The deviation of the $C$-$y$ relation from linearity for $y=0.13$ indicated that the effect of the additional magnetic interactions was pronounced at $y\ge0.1$.
Furthermore, as we will see, the peak width in the $\mathbf{Q}$ scan of magnetic scattering is actually smaller for $y=0.13$ than for $y=0.09$ (see Fig.~\ref{akane}).
As a result, crystals with $y=0.09$ were the main focus of $\mu$SR and neutron scattering measurements to estimate the inherent Cu-spin correlations in (Bi,Pb)2201.

At low $H$ and at low temperatures, a spin-glass phase is observable for all the Fe-doped samples.
As typically shown in Fig.~\ref{imp-dep}(b), a clear spin-glass transition arises in $\chi$ at $T_{\rm sg}=9$~K for $y=0.09$ when $H$ is applied along the $c$-axis.
The anisotropy between $\chi_{c}$ and $\chi_{ab}$ is attributed to the direction of the Fe spins,  although this has not been clarified as yet.

\subsection{Estimation of hole concentration}

The as-grown pristine sample is expected to be overdoped, due to the increase in $T_{\rm c}^{\rm on}$ by hole reduction.
The in-plane electrical resistivity $\rho_{ab}$, which was measured by a standard DC four-terminal method, indeed shows a normal metallic transport above  $T_{\rm c}^{\rm on}$ [Fig.~\ref{imp-dep}(c)].
Quantitatively, the doping rate of the pristine as-grown sample is determined to be $p = 0.28(2)$ from ARPES measurements.~\cite{ARPES}
This is close to the value (0.27) in Ref.~\onlinecite{maeda90} and fairly consistent with the universal dome-shaped superconducting phase diagram.~\cite{tallon95}

Moreover, $p$ of the 9\% Fe-doped sample was measured by ARPES, resulting in $p = 0.23(2)$~\cite{ARPES}.
This reduction in $p$ caused by Fe doping indicates a formation of Fe$^{3+}$ charge states, and it is consistent with preliminary measurements of $^{57}$Fe M\"{o}ssbauer spectroscopy,~\cite{kobayashi} and the edge energy of Fe-$K$ absorption in XAFS.~\cite{matsumura}
Fe doping of 9\% is sufficient for destroying the superconductivity in the heavily overdoped phase, and it causes an increase in the residual resistivity and carrier localization at low temperatures, as is evident from the upturn in $\rho_{ab}$ [Fig.~\ref{imp-dep}(c)].

 \begin{figure}[t]
  \centering
 \includegraphics[width=6.3cm]{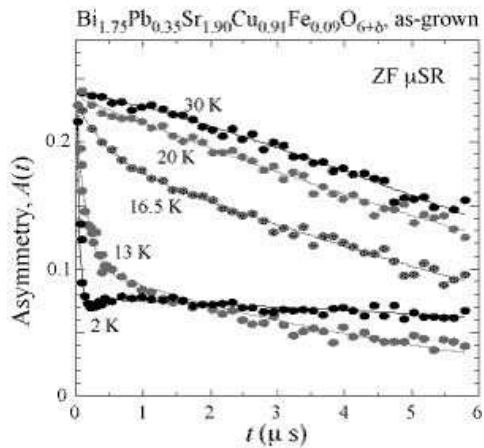}%
 \caption{ZF-$\mu$SR time spectra for the $y=0.09$ sample at low temperatures.
The single crystals were placed in a He gas-flow cryostat such that the CuO$_2$ plane faced the muon beam (i.e., muon spin polarization vector $\parallel$ $c$-axis of the sample).
$A(t)$ is given by $A(t) = \{ F(t)-\alpha B(t) \} / \{ F(t)+\alpha B(t) \}$, where $F(t)$ and $B(t)$ are the total muon events of a forward and a backward counter, respectively. 
$\alpha$ is a calibration factor.
The curved lines are fitted in the manner described in Ref.~\cite{watanabe02}.
 } 
\label{mSR}
 \end{figure}

\subsection{Zero-field $\mu$SR}

Zero-field (ZF)  $\mu$SR measurements using a positive muon beam at TRIUMF in Vancouver, Canada, were carried out at the M15 beamline.
Figure~\ref{mSR} shows the temperature variation of the asymmetry parameter $A(t)$ for the as-grown single crystals with $y=0.09$.
The Gaussian decay of the asymmetry found at 30~K and 20~K changes to a much more rapid relaxation at $T=$ 16.5~K and 13~K, which we attribute to the development of quasistatic electronic spin correlations.
At the base temperature of 2 K, a short-lived oscillatory signal is observed at $t$$<$1 $\mu \rm{s}$ and $A$($t$$>$1 $\mu \rm{s})$ remains nearly constant [$\sim$$A(0)/3$];
this indicates a static, short-range magnetic order over the entire sample.
The time spectra are analyzed by assuming three components in $A(t)$~\cite{watanabe02}; 
two of them indicate slow and fast depolarization, and the other expresses a muon spin precession.
The fast depolarization starts to appear at $T_{\mu{\rm SR}}$$\sim$20~K, 
and the internal field at the muon site is estimated to be $\sim$100~Oe at 2~K from the precession frequency.
For reference, these ZF-$\mu$SR measurements were also carried out using as-grown pristine crystals.
The nearly temperature-independent $A(t)$ below 30~K confirmed that  no static magnetic order is present down to 2~K, as in the case of the pristine (Bi,La)2201 crystal.~\cite{russo07}

\begin{figure}[t]
\centering
 \includegraphics[width=7.7cm]{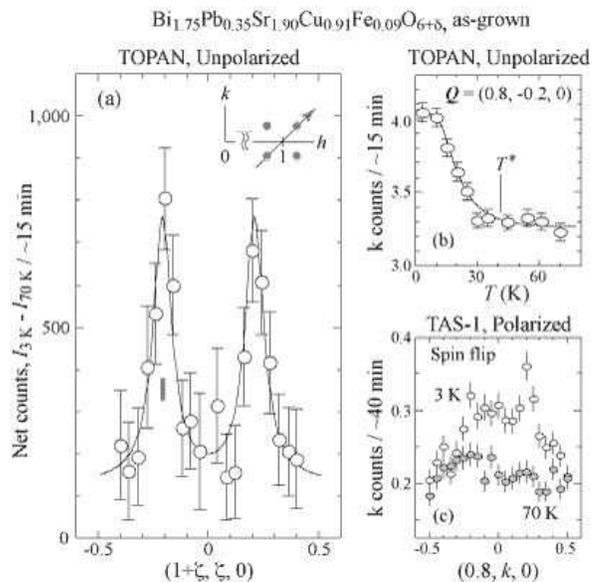}%
 \caption{Elastic scattering from the $y=0.09$ sample.
(a) Difference plot of $\textbf{Q}$ spectra obtained on TOPAN (neutron energy $E=14.7$~meV) using pyrolytic graphite (PG) $(002)$ reflections in the monochromator and analyzer.
The $q$ resolution is represented by the vertical bar.
Inset shows the scan trajectory parallel to the Cu-O-Cu bond axis.
(b) The thermal evolution of the incommensurate peak intensity.
The solid line is drawn as a guide.
(c) Temperature variation of $\textbf{Q}$ spectra observed using polarized neutrons at TAS-1 ($E=14.7$~meV)
in the spin-flip channel with a horizontal field of $\sim$20~Oe applied along the scattering vector.
Heusler $(111)$ reflections were used in the monochromator and analyzer.
Contaminations by higher-order incident neutrons were suppressed by inserting PG filters in the neutron beam path.
 } 
\label{neutron}
 \end{figure}

\subsection{Elastic neutron scattering}

Neutron scattering experiments were performed to elucidate the static spin correlations indicated by bulk magnetic susceptibility and $\mu$SR measurements.
The magnetic scattering was investigated by triple-axis neutron spectroscopy in the $(h,k,0)$ scattering plane, where $a^{*}$ and $b^{*}$ are $\sim$1.18 and 1.17~\AA$^{-1}$, respectively.
Unpolarized (polarized) neutron-scattering experiments were performed on triple-axis spectrometers TOPAN and AKANE (TAS-1) at the research reactor JRR-3 of the Japan Atomic Energy Agency in Tokai, Japan.
Figure~\ref{neutron}(a) shows that peaks due to distinct incommensurate elastic scattering 
occur near $(1,0,0)$ at low temperatures for $y=0.09$.
The diffuse peaks, which start appearing below $T^{*}$$\sim$ 40~K [Fig.~\ref{neutron}(b)], correspond to antiferromagnetic short-range modulations propagating along the Cu-O-Cu bond axes.
The direction of the spin modulation is identical to that observed for the superconducting La214 and Y123 systems. 
Further, results of the polarized-neutron analysis performed in a spin-flip channel confirm that the diffuse incommensurate peaks appearing below $T^{*}$ are of magnetic origin, as shown in Fig.~\ref{neutron}(c).
By assuming four Lorentzian peaks at $\mathbf{Q}=(1+\delta, \pm\delta, 0)$ and $(1-\delta, \pm\delta, 0)$ with a HWHM $\kappa$, $\delta=0.21 (1)$ and $\kappa=0.074 (13)$~\AA$^{-1}$ are extracted by resolution-convoluted fitting to the difference plot shown in Fig.~\ref{neutron}(a).
The magnetic modulation period and magnetic correlation length $\xi (=1/\kappa)$ are found to be $\sim$$4.8a_{\rm tet}$ and $\sim$$3.5a_{\rm tet}$, respectively, where $a_{\rm tet}\approx a/\sqrt{2}\sim3.8$~\AA.

\begin{figure}[t]
\centering
 \includegraphics[width=5.8cm]{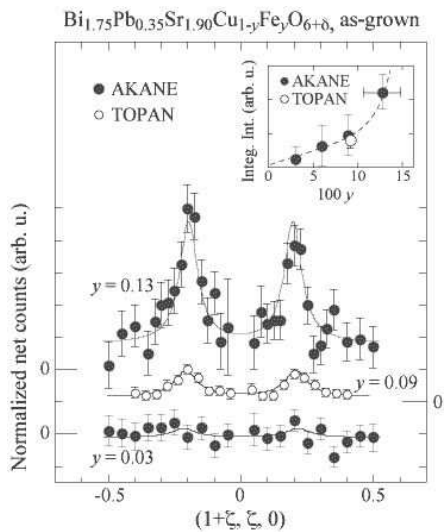}%
 \caption{
Difference plots for $y=0.13, 0.09,$ and $0.03$ measured at AKANE ($E=19.4$~meV) and TOPAN. 
The neutron energy  was selected on AKANE using a Ge $(311)$ monochromator and a PG $(002)$ analyzer.
The Fe-doped single crystals were approximately 30$\times$5$\times$2~mm$^{3}$ in size, and $2\eta<0.5^{\circ}$ for $y\le 0.09$ and $\sim$$1^{\circ}$ for $y=0.13$ in mosaicness.
The relative intensity spectra are plotted with offset. 
The curved lines are fitted to a pair of Lorentzians and a constant.
Dependence of the $q$-integrated intensity on $y$ is shown in the inset.
 } 
\label{akane}
 \end{figure}

As shown in Fig.~\ref{akane}, $\delta$ changes little upon Fe  doping within this concentration range, but the peak width at $y=0.13$ becomes smaller than that at $y=0.09$.
The normalized $q$-integrated intensity presented in the inset shows a linear increase up to $y=0.09$ but an additional increase at $y=0.13$, thereby indicating a similarity with the non-linear dependence of $C$ on $y$.

\section{Discussions}

\subsection{Spin clusters around Fe}

Summarizing the data, we can obtain an overall picture of the Fe-spin-induced ISCs in overdoped (Bi,Pb)2201.
Since the average Fe-Fe separation ($\sim$$a_{\rm tet}/\sqrt{y})$ nearly corresponds to $\xi$ when $y=0.09$, small clusters with diameter $\sim\xi$ are formed around Fe. 
The linear dependences of both $C$ and the neutron scattering intensity on $y$ below $y=0.09$, together with the weak dependence of $\delta$ on $y$, suggest that the clusters do not interact much with each other. 
However, Fe doping beyond $y=0.1$ may introduce additional effects of Fe-Fe and/or inter cluster interactions,  causing the dependence of $C$ and scattering intensity on $y$ to deviate from linearity.  

The onset temperature of the induced ISCs depends on the nature of probe; $T^{*}$$\sim$ 40~K determined by neutron scattering is considerably higher than $T_{\rm sg}$ and $T_{\mu{\rm SR}}$ determined by magnetization and $\mu$SR measurements, respectively. 
Therefore, the induced ISCs are quasistatic in nature similar to the case of the spin-glass phase of La214~\cite{wakimoto00}.
Our recent reinvestigation of magnetic susceptibility found that $\chi_{ab}(T)$ of Fig.~\ref{imp-dep}(a) starts to deviate from the Curie-Weiss law below $\sim T^{*}$.~\cite{wakimotoH}
This phenomena is also seen in La214 and it might be a precursor of the spin-glass transition.~\cite{wakimoto00}

Although the local charge state of Fe is most likely Fe$^{3+}$, the obtained $p_{\rm eff}$ is much smaller than expected for high-spin $S=5/2$ ($p_{\rm eff}=5.9$), but similar to that in Fe-doped Bi2212 ($p_{\rm eff}=3.6$)~\cite{maeda90-imp}.
To explain the reduced $p_{\rm eff}$, we speculate that the Fe spin strongly couples with neighbor Cu spins and/or ligand hole spins in an anti-parallel fashion.
Note that a similar reduction in the effective spin value is observed in Ni-doped La214.~\cite{hiraka09}

\begin{figure}[t]
\centering
 \includegraphics[width=7cm]{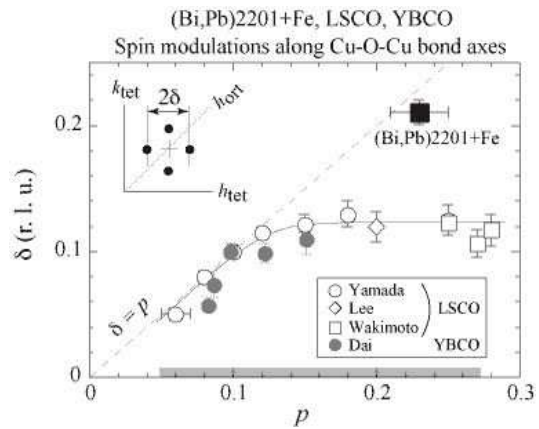}%
 \caption{
Comparison of $\delta$ between Fe-doped (Bi,Pb)2201, pristine La214~\cite{yamada98,lee03,wakimoto04}, and Y123~\cite{dai01}.
Dynamical data are plotted for La214 ($\omega=3$--6~meV at $T\sim T_{\rm c}$) and Y123 ($\omega\ll$ (resonance energy) at $T<T_{\rm c}$).
The $p$ range of the superconducting phase is indicated by a thick bar at the base line.
 The solid line is a guide for La214 and Y123 data.
  } 
\label{incom}
 \end{figure}

\subsection{ISCs in overdoped (Bi,Pb)2201}

The highlight of the current study is the observation of quasi-static ISCs induced by Fe dopants in overdoped (Bi,Pb)2201.
To the best of our knowledge, this is the first observation of ISCs in the (Bi,Pb)2201 system, although the aid of large Fe spins was needed to see them.
Hence, our finding supports the idea that the ISCs are a common feature in high-$T_{\rm c}$ cuprate superconductors.
At the same time, we can provide neutron scattering data for the overdoped region of a system other than La214.
The ISCs in overdoped (Bi,Pb)2201 possess the largest value of $\delta$ among the high-$T_{\rm c}$ cuprates studied so far.
Figure~\ref{incom} shows $\delta$, determined from low-energy modulations along the Cu-O-Cu bond axes in La214~\cite{yamada98,lee03,wakimoto04} and Y123.~\cite{dai01}
Intriguingly, $\delta$ of the Fe-doped (Bi,Pb)2201 studied here appears to follow an extrapolation of the linear relationship $\delta=p$ established in La214 for $p\alt1/8$.

One conjecture for this behavior is that the linearity manifests a gradual increase in density of dynamic stripes against $p$ in pristine overdoped (Bi,Pb)2201.
In this picture, the Fe-spin injection may contribute to carrier localization [Fig.~\ref{imp-dep}(c)], reducing the characteristic frequency of dynamic ISCs locally, pinning stripe correlations around Fe sites, and thus resulting in the spin clusters.
This is essentially the same story as has been proposed for the impact of Zn doping on underdoped La214\cite{hirota01}; however, there are some problems in applying it to the overdoped regime.  For one thing, the Zn dopants that cause pinning of stripes in La214 also cause a reduction in the spin correlation length, in contrast to the present behavior.  Even more significant is the question of whether
spin stripes (or at least spin stripes induced by charge stripes) with a period as short as $5 a_{\rm tet}$ could be energetically stable, even locally.
To further test the possibility of Fe-induced stripes and to compare with stripes in La214, an investigation of magnetic-field effects on neutron scattering and electric resistivity in Fe-doped (Bi,Pb)2201 is in progress~\cite{wakimotoH}.

An alternative possibility is that the magnetic correlations are determined by the electronic Fermi surface.  A simple Fermi-surface nesting scenario, as in Cr, seems unlikely as one would not expect nesting behavior to be established or amplified by magnetic impurity ions, which should cause considerable scattering of quasiparticles [as indicated by the increased in-plane resistivity, Fig.~\ref{imp-dep}(c)]. On the other hand, the presence of the Fe ions does suppress the superconductivity and presumably closes the superconducting gap, which would increase the density of states available for magnetic interactions.  We note that a crossover in the electronic response to magnetic impurities between under- and over-doped regimes has been observed in a recent study of Ni-doped La214.\cite{tanabe10}   Perhaps the magnetism involves an RKKY-type of coupling between Fe moments via the conduction electrons, as in dilute Cu-Mn alloys.\cite{lamelas95}  Such an analogy would encompass both the spin-glass behavior and the ISCs.   In any case, more work is required to understand the nature of the (induced) magnetism in overdoped (Bi,Pb)2201.

\section{Summary}
Quasi-static spin correlations induced by Fe dopants in overdoped  (Bi,Pb)2201 were studied by magnetization, $\mu$SR, and neutron-scattering measurements through Fe doping.  The magnetization measurements indicate spin-glass-like ordering below 10~K for a crystal with $y=0.09$, and the $\mu$SR measurements on the same sample confirm the presence of static magnetic order at 2~K.  
ISCs propagating along the Cu-O-Cu bonding axes as in the superconducting phase of other high-$T_{\rm c}$ cuprates are detected by elastic neutron scattering at temperatures below 40~K, with the intensity saturating below 10~K.  
The large value of $\delta$ ($\sim$0.2) as well as the coincidence with $p$ are unexpected and quite different from the saturation of $\delta$ in overdoped La214.
An explanation in terms of stripes seems unlikely.  An alternative possibility involves the coupling between Fe moments through conduction electrons.  Further experimental work is required in order to come up with a proper understanding of the dopant-induced static magnetism in overdoped (Bi,Pb)2201.\\

\vspace{10mm}

\begin{acknowledgments}
We are grateful to K.~Kudo, H.~Kobayashi, D.~Matsumura, and T.~Sato for their helpful discussions. 
We also thank K.~Nemoto and M.~Sakurai for their assistance in the neutron-scattering and crystal-growth experiments, respectively.
This study was carried out under the Common-Use Facility Program of JAEA, and the Quantum Beam Technology Program of JST.
The study performed at Tohoku University was supported by a Grant-In-Aid for Science Research C (19540358) and B (19340090) from the MEXT.
JMT is supported by the U.S. Department of Energy, Office of Basic Energy Sciences, Division of Materials Sciences and Engineering, under Contract No.~DE-AC02-98CH110886.
\end{acknowledgments}

\bibliography{basename of .bib file}

\end{document}